\documentclass[aps, prl, reprint, amsmath,amssymb, superscriptaddress]{revtex4-1}

\usepackage{graphicx}
\usepackage{xcolor,color}

\usepackage{dcolumn}
\usepackage{bm}
\usepackage{relsize}

\usepackage[normalem]{ulem} 

\newcommand{\BSCCO}{Bi$_2$Sr$_2$CaCu$_2$O$_{8+\delta}$ }
\newcommand{\BSCO}{Bi$_2$Sr$_{2-x}$La$_x$CuO$_{6+\delta}$ }

\begin{document}

\title{Ubiquitous suppression of the nodal coherent spectral weight in Bi-based cuprates}

\author{M.\,Zonno}
\email[]{mzonno@phas.ubc.ca}
\thanks{Equally contributed author}
\affiliation{Quantum Matter Institute, University of British Columbia, Vancouver, BC V6T 1Z4, Canada}
\affiliation{Department of Physics $\&$ Astronomy, University of British Columbia, Vancouver, BC V6T 1Z1, Canada}
\author{F.\,Boschini}
\thanks{Equally contributed author}
\affiliation{Quantum Matter Institute, University of British Columbia, Vancouver, BC V6T 1Z4, Canada}
\affiliation{Department of Physics $\&$ Astronomy, University of British Columbia, Vancouver, BC V6T 1Z1, Canada}
\author{E.\,Razzoli}
\affiliation{Quantum Matter Institute, University of British Columbia, Vancouver, BC V6T 1Z4, Canada}
\affiliation{Department of Physics $\&$ Astronomy, University of British Columbia, Vancouver, BC V6T 1Z1, Canada}
\affiliation{SwissFEL, Paul Scherrer Institut, 5232 Villigen PSI, Switzerland}
\author{M.\,Michiardi}
\affiliation{Quantum Matter Institute, University of British Columbia, Vancouver, BC V6T 1Z4, Canada}
\affiliation{Department of Physics $\&$ Astronomy, University of British Columbia, Vancouver, BC V6T 1Z1, Canada}
\affiliation{Max Planck Institute for Chemical Physics of Solids, N{\"o}thnitzer Stra{\ss}e 40, Dresden 01187, Germany}
\author{M.\,X.\,Na}
\affiliation{Quantum Matter Institute, University of British Columbia, Vancouver, BC V6T 1Z4, Canada}
\affiliation{Department of Physics $\&$ Astronomy, University of British Columbia, Vancouver, BC V6T 1Z1, Canada}
\author{S.\,Dufresne}
\affiliation{Quantum Matter Institute, University of British Columbia, Vancouver, BC V6T 1Z4, Canada}
\affiliation{Department of Physics $\&$ Astronomy, University of British Columbia, Vancouver, BC V6T 1Z1, Canada}
\author{T.\,M.\,Pedersen}
\author{S.\,Gorovikov}
\affiliation{Canadian Light Source, Inc.44 Innovation Boulevard, Saskatoon, SK S7N 2V3, Canada}
\author{S.\,Gonzalez}
\author{G.\,Di\,Santo}
\author{L.\,Petaccia}
\affiliation{Elettra Sincrotrone Trieste, Strada Statale 14 km 163.5, 34149 Trieste, Italy}
\author{M.\,Schneider}
\author{D.\,Wong}
\author{P.\,Dosanjh}
\affiliation{Quantum Matter Institute, University of British Columbia, Vancouver, BC V6T 1Z4, Canada}
\author{Y.\,Yoshida}
\author{H.\,Eisaki}
\affiliation{National Institute of Advanced Industrial Science and Technology (AIST), Tsukuba 305-8568, Japan}
\author{R.\,D.\,Zhong}
\affiliation{\mbox{Condensed Matter Physics and Materials Science, Brookhaven National Laboratory, Upton, NY, USA}}
\author{J.\,Schneeloch}
\affiliation{\mbox{Condensed Matter Physics and Materials Science, Brookhaven National Laboratory, Upton, NY, USA}}
\author{G.\,D.\,Gu}
\affiliation{\mbox{Condensed Matter Physics and Materials Science, Brookhaven National Laboratory, Upton, NY, USA}}
\author{A.\,K.\,Mills}
\author{S.\,Zhdanovich}
\author{G.\,Levy}
\author{D.\,J.\,Jones}
\affiliation{Quantum Matter Institute, University of British Columbia, Vancouver, BC V6T 1Z4, Canada}
\affiliation{Department of Physics $\&$ Astronomy, University of British Columbia, Vancouver, BC V6T 1Z1, Canada}
\author{A.\,Damascelli}
\email[]{damascelli@physics.ubc.ca}
\affiliation{Quantum Matter Institute, University of British Columbia, Vancouver, BC V6T 1Z4, Canada}
\affiliation{Department of Physics $\&$ Astronomy, University of British Columbia, Vancouver, BC V6T 1Z1, Canada}

\begin{abstract}
High-temperature superconducting cuprates exhibit an intriguing phenomenology for the low-energy elementary excitations. In particular, an unconventional temperature dependence of the coherent spectral weight (CSW) has been observed in the superconducting phase by angle-resolved photoemission spectroscopy (ARPES), both at the antinode where the d-wave paring gap is maximum, as well as along the gapless nodal direction. Here, we combine equilibrium and time-resolved ARPES to track the temperature dependent meltdown of the nodal CSW in Bi-based cuprates with unprecedented sensitivity. We find the nodal suppression of CSW upon increasing temperature to be ubiquitous across single- and double-layer Bi cuprates, and uncorrelated to superconducting and pseudogap onset temperatures. We quantitatively model both the lineshape of the nodal spectral features and the anomalous suppression of CSW within the Fermi-Liquid framework, establishing the key role played by the normal state electrodynamics in the description of nodal quasiparticles in superconducting cuprates.
\end{abstract}

\maketitle

Copper-oxide high-T$_c$ superconductors host a variety of emerging and competing quantum phases. On the one hand, their low temperature physics is mainly characterized by unconventional d-wave superconductivity (SC); on the other hand, the normal state is actually a ``strange metal" exhibiting various complex phenomena, such as pseudogap physics and charge density wave order \cite{damascelli2003angle, comin2016resonant, keimer2015quantum, frano2019charge}. 
Since the discovery of cuprates, the existence (or lack thereof) of well-defined quasiparticles in the normal state, as well as their temperature and doping dependence, have been the subject of intense debate. The characteristic onset temperature T$_c$ has been associated not only with the SC phase transition but also with the emergence of coherent quasiparticles \cite{sawatzky1989testing, shen1999novel}. In this regard, specific insight has come from angle-resolved photoemission spectroscopy (ARPES), either at equilibrium or in pump-probe configuration, which provides access to the low-energy electronic structure and its dynamics
\cite{sawatzky1989testing, ding1996angle, shen1999novel, feng2000signature, ding2001coherent, lanzara2001evidence, johnson2001doping, fournier2010loss, graf2011nodal, smallwood2014time, Boschini2018}. 
In particular, an unconventional temperature dependence of the coherent spectral weight (CSW) for antinodal quasiparticles has been reported by ARPES studies on \BSCCO (Bi2212); this was described either as a manifestation of the c-axis superfluid density, or as a testament of the different role played by the quasiparticle coherence in the emergence of SC in underdoped versus overdoped regimes \cite{feng2000signature, ding2001coherent}. 

Towards unveiling the detailed evolution of quasiparticle weight and lifetime from normal to SC state, the study of the antinodal region of cuprates is hindered by the interplay of several intertwined contributions, such as SC gap, pseudogap, charge order, as well as strong band renormalization. More promising is in principle the exploration of the gapless and more pristine nodal direction; in this regard, similarly to what was reported at the antinode, a time-resolved ARPES (TR-ARPES) study of optimally doped Bi2212 has suggested a direct relation between nodal CSW and the SC condensate \cite{graf2011nodal}. However, a recent theoretical work has questioned such a direct link, proposing instead the key role of a different competing order \cite{zheng2017coherent}. As a consequence, for either nodal and antinodal regions of the CuO$_2$ plane electronic structure, a comprehensive and conclusive description of the evolution of CSW across the superconducting-to-normal-state phase transition is still missing.

In laying the basis for discussing the T-dependence of CSW in cuprates, we remark that the many-body renormalization of lifetime and pole structure is captured by the complex electron self energy $\Sigma(\omega, \text{T})$=$\Sigma^{'}(\omega, \text{T})+i\, \Sigma^{''}(\omega, \text{T})$, where the momentum dependence is commonly neglected. ARPES experiments provide access to $\Sigma(\omega, \text{T})$ by measuring the single-particle removal spectral function $A(\mathbf{k},\omega,\text{T})$, with its corresponding many-body self-energy corrections \cite{damascelli2003angle}. 
While in the hypothetical scenario of a non-interacting system $A(\mathbf{k},\omega,\text{T})$ collapses to a series of delta functions whose area is not expected to vary as a function of temperature, in the presence of electron interactions $\Sigma(\omega,\text{T})$ may lead to the broadening and redistribution of the spectral weight within the momentum-energy phase space. Therefore, one may argue that the observed T-dependence of CSW for a well defined momentum -- specifically the Fermi momentum $\mathbf{k}_\mathrm{F}$, CSW($\mathbf{k}_\mathrm{F}$,\,T)=$\int_{-\infty}^{\infty}A(\mathbf{k}_\mathrm{F},\omega,\text{T}) \, d \omega$ -- is determined by the underlying electron interactions encoded in $\Sigma(\omega,\text{T})$.

In this Letter, we establish the role played by the normal-state self energy to the integrity of quasiparticles by tracking the T-dependence of both the $\omega$=0 and energy-integrated nodal spectral function at $\mathbf{k}_\mathrm{F}$. 
This is achieved -- for both single- and bi-layer Bi cuprates at various dopings -- by employing TR-ARPES, which offers an enhanced sampling and signal-to-noise for the T-dependence of the nodal spectra. 
This approach, corroborated by conventional equilibrium ARPES, reveals a ubiquitous suppression of nodal CSW that bears no relation to T$_c$, but instead directly stems from the $\omega$- and T-dependence of the complex self energy. By analyzing the data in terms of Fermi Liquid (FL) \cite{pines1966theory} and Marginal Fermi Liquid (MFL) \cite{varma1989phenomenology} models, we find that solely the FL framework reproduces comprehensively the T-dependence of the nodal CSW, as well as the overall nodal quasiparticle phenomenology encoded in the lineshape and amplitude of energy and momentum distribution curves (EDCs and MDCs).

\begin{figure}[t]
\centering
\includegraphics[scale=1.03]{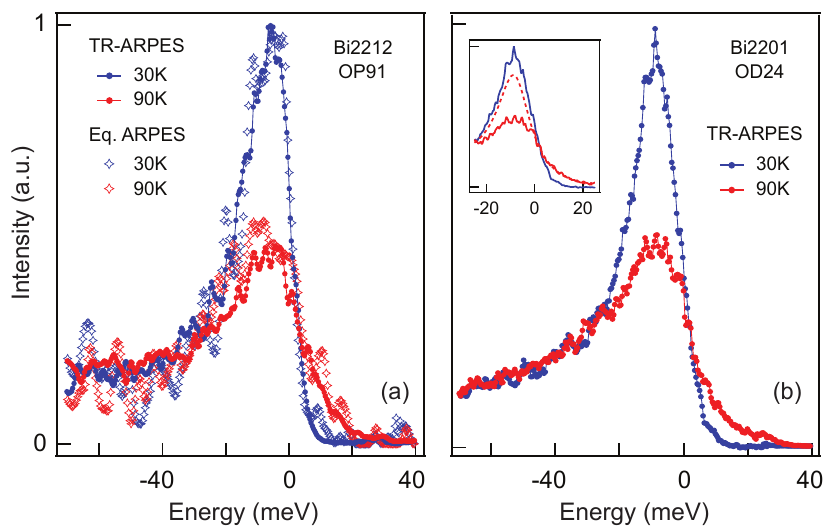}
\caption[Fig1]{(a) Energy distribution curves (EDCs) at the Fermi momentum $\mathbf{k}_\mathrm{F}$ along the nodal direction of Bi2212-OP91 for 30\,K and 90\,K, as measured by TR-ARPES ($h\nu$=$6.2$\,eV) and equilibrium ARPES ($h\nu$=$27$\,eV). (b) Nodal EDCs at $\mathbf{k}_\mathrm{F}$ for Bi2201-OD24 acquired by TR-ARPES. Inset: comparison between the same experimental data (solid lines) and the EDC obtained from the 30\,K data via the thermal broadening of the Fermi-Dirac distribution at 90\,K (dashed line). All EDCs have been deconvoluted from the energy resolution broadening via the Lucy-Richardson algorithm \cite{Boschini2018,Yang2008deconv,Razzoli2013NodelessGap} (TR-ARPES energy resolution is 11\,meV and 18\,meV for panels (a) and (b), respectively; equilibrium ARPES energy resolution is 5.3\,meV).}
\label{Fig1} 
\end{figure}

\BSCO (Bi2201) and \BSCCO (Bi2212) samples have been aligned via Laue diffraction along the nodal $\overline{\Gamma}$--$\overline{\text{Y}}$ direction in order to avoid replica bands \cite{king2011structural, saini1997topology, yang1998crystal}, and cleaved in vacuum at pressures and base temperature lower than $8\cdot10^{-11}$\,Torr and $20\,$K, respectively. Equilibrium ARPES measurements have been performed at the Quantum Materials Spectroscopy Centre (QMSC) beamline at the Canadian Light Source using 27\,eV $\sigma$-polarized light, with energy and momentum resolutions better than 5.5\,meV and 0.007\,$\text{\AA}^{-1}$, respectively. Additional ARPES experiments have been conducted at the BaDElPh endstation at the Elettra synchrotron with tunable photon energy in the range $7$\,-\,$30\,$eV. TR-ARPES data have been acquired at the UBC-Moore Center for Ultrafast Quantum Matter by pumping and probing with $\sigma$-polarized 1.55\,eV and 6.2\,eV pulses, and with overall energy, momentum, and temporal resolution of 18\,meV, 0.0025\,$\text{\AA}^{-1}$, and 250\,fs, respectively \cite{Boschini2018}, unless specified differently in the text. 

In this work we employ TR-ARPES as a tool to finely tune the transient electronic temperature via optical pumping, and relate it to changes of the nodal spectral features \cite{giannetti2016ultrafast, graf2011nodal, boschini2020emergence}. Since each pump-probe delay is acquired in the same experimental conditions, this dynamical approach naturally sets the same background baseline for all EDCs and MDCs, allowing for a direct measure of the relative variation of the spectral weight as a function of the effective electronic temperature.
Figure\,\ref{Fig1} shows EDCs measured by TR-ARPES at $\mathbf{k}_\mathrm{F}$ along the nodal direction, for two different temperatures of optimally-doped Bi2212 (T$_c$=$91\,$K, OP91) and overdoped Bi2201 (T$_c$=$24\,$K, OD24) -- energy resolution broadening was removed via the Lucy-Richardson algorithm \cite{Boschini2018,Yang2008deconv,Razzoli2013NodelessGap}. For both compounds, we observe a clear suppression of CSW close to the Fermi energy (E$_\mathrm{F}$), as the temperature increases from 30 to 90\,K. Note that such suppression cannot be ascribed to a transient state induced by the pump excitation, as evidenced in Fig.\,\ref{Fig1}a by the agreement between the nodal EDCs probed via TR-ARPES at 6.2\,eV and conventional equilibrium ARPES at 27\,eV. This observation not only indicates that the suppression of CSW is inherently related to the temperature of the system, thus validating the TR-ARPES approach as our main experimental strategy throughout the entire work, but excludes the explicit breaking of the sudden-approximation with low-energy photons \cite{DessauSudden_PRL2016} (also note that only the antibonding state is probed using both 6.2\,eV and 27\,eV, and nonlinear effects in the photoelectrons detection \cite{Reber2014nonlinearity} have been ruled out; details in Supplementary Materials). Finally, as illustrated in the inset of Fig.\,\ref{Fig1}b, the mere thermal broadening of the Fermi-Dirac distribution at 90\,K does not account for the observed T-dependence of the experimental data.

\begin{figure}[t]
\centering
\includegraphics[scale=1.01]{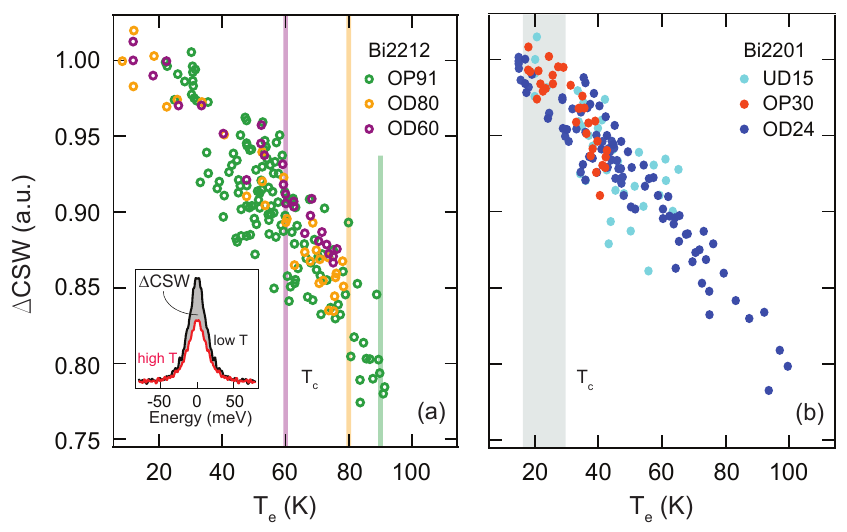}
\caption[Fig2]{(a) Relative variation of the nodal coherent spectral weight $\Delta$CSW as a function of the electronic temperature T$_\mathrm{e}$ tracked via TR-ARPES, for three different doping levels of Bi2212. The inset illustrates the difference in the integrated area of the symmetrized EDCs in the $[-0.08, 0.08]\,$eV range, which defines $\Delta$CSW. (b) Same as in (a), but for three doping levels of Bi2201. Colored (a) and gray (b) vertical bars mark the range of T$_c$ for the different compounds.}
\label{Fig2} 
\end{figure}

To quantitatively describe the variation of CSW as a function of the electronic temperature (T$_\mathrm{e}$), we focus our analysis on the evaluation of the area of the symmetrized TR-ARPES EDCs (SEDCs) at $\mathbf{k}_\mathrm{F}$. In fact, provided the particle-hole symmetry of the spectral function -- which has been experimentally verified at $\mathbf{k}_\mathrm{F}$ in the near-nodal region of Bi2212 \cite{Boschini2018, matsui2003bcs} -- SEDCs directly map into the spectral function, SEDC($\omega$) $\propto$ A($\mathbf{k}_\mathrm{F}, \omega$) \cite{norman1998destruction, parham2017ultrafast}. Here, we define the relative suppression of coherent spectral weight, $\Delta$CSW, as the difference between the integrated area under the SEDCs at $\mathbf{k}_\mathrm{F}$ at low and high T$_\mathrm{e}$ (\textit{i.e.} before and after the pump excitation), in the $[-0.08, 0.08]\,$eV range (see inset of Fig.\,\ref{Fig2}a). This experimental strategy makes our analysis bias-free from the choice of specific models or fitting procedures of the SEDCs. The obtained $\Delta$CSW is plotted directly as a function of T$_\mathrm{e}$ in Fig.\,\ref{Fig2} for different doping levels of Bi2212 and Bi2201. Despite the large number of dopings and related T$_c$ across the studied compounds, a remarkable similarity characterizes the T-dependent suppression of $\Delta$CSW(T$_\mathrm{e}$) in the explored 100\,K temperature range. Most important, the decrease of $\Delta$CSW(T$_\mathrm{e}$) does not exhibit any abrupt modification across the superconducting-to-normal state phase transitions, and continues for T$_\mathrm{e}\gg$T$_c$ (colored and gray bars in Fig.\,\ref{Fig2} highlight T$_c$ for the various dopings studied here). These findings demonstrate the absence of a direct contribution of the superconducting phase to $\Delta$CSW(T$_\mathrm{e}$), and establish the T-dependence of the nodal coherent spectral weight as a ubiquitous behavior of Bi-based cuprates. The data reported in Fig.\,\ref{Fig2} also exclude a possible relation of the observed nodal suppression of CSW to the pseudogap phenomenon. In fact, despite no equilibrium pseudogap having been reported for the very overdoped Bi2212-OD60 compound \cite{kawasaki2010carrier, vishik2012phase, chatterjee2011electronic}, its $\Delta$CSW (purple open circles in Fig.\,\ref{Fig2}a) is comparable to that observed for the other compounds.

\begin{figure*}[]
\centering
\includegraphics[scale=1.045]{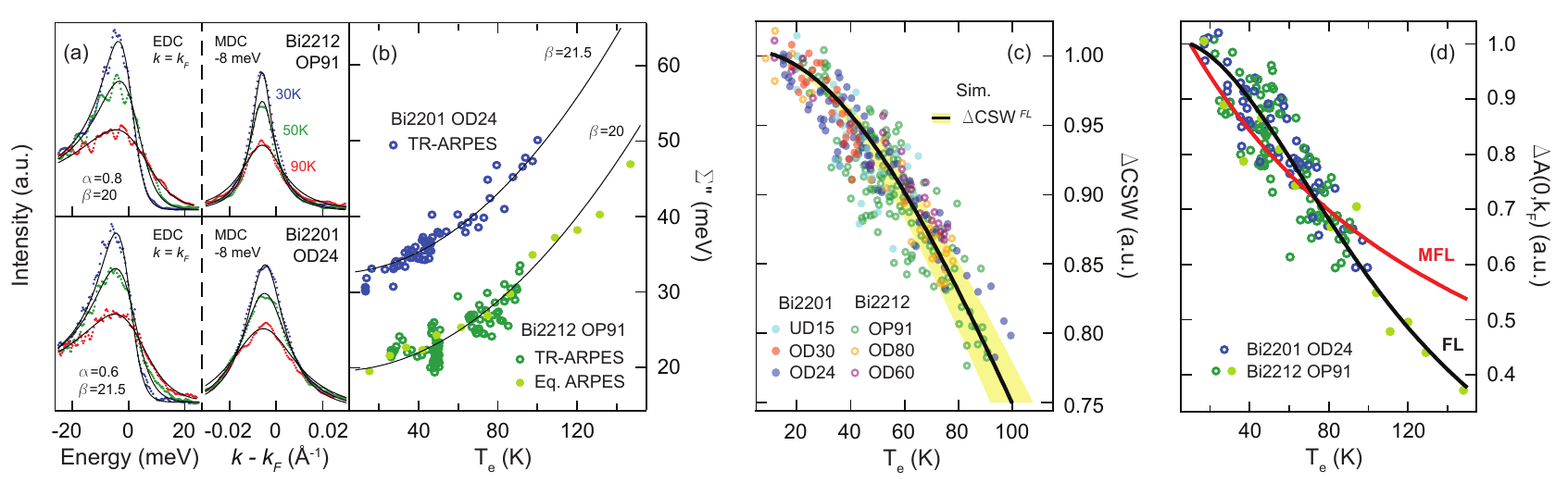}
\caption[Fig3]{(a) Nodal EDCs at $\mathbf{k}_\mathrm{F}$ and MDCs at $-8$\,meV for Bi2212-OP91 and Bi2201-OD24, probed by TR-ARPES at different temperatures. Black solid lines represent the results of the EDC-MDC global fitting procedure using Eq.\,\ref{eq:2}, as explained in the main text. Optimal fits to the data are found with $(\alpha$=0.8,\, $\beta$=20) and $(\alpha$=0.6,\, $\beta$=21.5) for Bi2212-OP91 and Bi2201-OD24, respectively. (b) Temperature evolution of the imaginary part of the electron self energy extracted via fitting of MDCs at $-8$\,meV (integration windows $\pm\,5$meV) via a Lorentzian-like function for Bi2201-OD24 and Bi2212-OP91. The solid black lines are quadratic curves defined by the $\beta$ coefficients obtained via the global fit procedure of panel (a). 
(c) Comparison between the experimental CSW suppression (data from Fig.\ref{Fig2}) and the simulated $\Delta$CSW$^{FL}$ within the FL model via Eq.\,\ref{eq:2}. The solid black line is computed for $\alpha$=0.7, $\beta$=20.75, $\Gamma$=0.02, while the yellow shading accounts for the parameters' range defined by the global fit. 
(d) Normalized temperature evolution of the spectral function $\Delta A(0,\mathbf{k}_\mathrm{F})$, for Bi2201-OD24 (TR-ARPES data) and Bi2212-OP91 (TR- and equilibrium ARPES). The integration range in momentum and energy is $0.005\,$\AA$^{-1}$ and $6\,$meV, respectively. The solid lines are best fits to the experimental data by using Eq.\,\ref{eq:3} for FL (black; $\Gamma$=0.02 and $\beta$=20.35) and MFL (red; $\Gamma$=0.019 and $\lambda$=0.96) models. All spectra have been deconvoluted from the energy resolution before extracting EDCs and MDCs in panels (a), (b), and (d).}
\label{Fig3} 
\end{figure*}

Having excluded any dependence on the characteristic T$_c$, we now explore the possible contribution of normal state properties to the T-dependence of CSW. In the following we demonstrate that the progressive meltdown of the nodal CSW can be understood in terms of the intrinsic $\omega$- and T-dependence of the self energy within the Fermi Liquid model. The FL self energy can be written as \cite{pines1966theory, damascelli2003angle}:
\begin{equation} \label{eq:2}
\Sigma _{FL} (\omega, \text{T})  = -\alpha \omega -i [ \Gamma +\beta\, (\omega^2 + \pi^2 \text{T}^2)] \, ,
\end{equation}
where $\alpha$, $\beta$, and $\Gamma$ are positive parameters, and $\Gamma$ represents a scattering rate term associated with static impurities, and thus independent of energy and temperature \cite{abrahams2000angle}. 
Using Eq.\,\ref{eq:2}, we can calculate the T-dependence of the spectral function and compute the resulting $\Delta$CSW$^{FL}$. We reiterate that the discussion is limited to $\mathbf{k}$=$\mathbf{k}_\mathrm{F}$ (\textit{i.e.} $\varepsilon ^ b_{\mathbf{k}}$\,=\,0), which allows for a direct comparison to the experimental data presented in Fig.\,\ref{Fig2}. In order to estimate the values of $\alpha$ and $\beta$, and thus provide a more quantitative modeling of $\Delta$CSW$^{FL}$, we developed a global fit analysis of EDCs and MDCs in terms of Eq.\,\ref{eq:2}. Figure\,\ref{Fig3}a shows the results of this global fitting procedure for three different temperatures of Bi2212-OP91 and Bi2201-OD24. We assume a bare velocity $v_0$=$3.8\,(3.35)\,$eV$\cdot\,\text{\AA}$ for Bi2212 (Bi2201), as reported from previous studies \cite{kaminski2005momentum,kordyuk2005bare, meevasana2008extracting}. Note that the parameter $\alpha$ can be related to the bare velocity by the expression $\alpha=\frac{v_0}{v_\mathrm{F}}-1$, where $v_\mathrm{F}$ is the renormalized Fermi velocity. Despite several mechanisms playing a role in determining the self energy, in our analysis we do not differentiate the various contributions, and assume a linear renormalized dispersion in the first 0.2\,eV below E$_\mathrm{F}$.
The best simultaneous global fit to EDCs and MDCs is achieved for $\alpha$=0.8 and $\beta$=20 $(\alpha$=0.6 and $\beta$=21.5) for Bi2212-OP91 (Bi2201-OD24). These values are consistent with the band renormalization reported in previous ARPES studies \cite{kordyuk2005bare, zhou2003universal, zhang2014ultrafast}, and describe well the quadratic temperature evolution of the imaginary part of the electron self energy (see Fig.\,\ref{Fig3}b), thus validating our global fit results.
We emphasize that the global fit procedure was tested against a general functional form of the imaginary part of the self energy, namely $\Sigma^{''}(\omega, \text{T}) = -\Gamma -\beta\, (\omega^2 + \pi^2 \text{T}^2)^{1/a}$. While $a$\,=\,1 describes the FL self energy, $a$\,=\,2 resembles the linear $\omega$- and T-dependence of the phenomenological MFL self energy \cite{varma1989phenomenology, abrahams2000angle}. The global fit analysis consistently converges to $a$\,=\,1\,$\pm$\,0.1, pointing towards a description of the experimental data in terms of the FL model.

Having established an effective range for the parameters in Eq.\,\ref{eq:2}, we now simulate the suppression of spectral weight $\Delta$CSW$^{FL}$ at $\mathbf{k}_\mathrm{F}$ (integration window $[-0.08, 0.08]\,$eV, normalized to the 10\,K value). This is plotted in Fig.\,\ref{Fig3}c superimposed to all the TR-ARPES experimental data shown in Fig.\,\ref{Fig2}. In particular, $\alpha$=0.7, $\beta$=20.75, and $\Gamma$=0.02 were used to simulate the solid black line, while the yellow shading corresponds to the parameters' range defined by the global fit in Fig.\,\ref{Fig3}a. A remarkable agreement is observed between the FL simulation and the experimental suppression of CSW, both in terms of functional form and magnitude, with a quenching of the CSW as large as $\sim\,25\%$ at 100\,K.
Despite the significant suppression of CSW at $\mathbf{k}_\mathrm{F}$, we note that the tomographic density of states along the nodal direction \cite{TDOSNatPhys} (\textit{i.e.} $\int_{-\infty}^{\infty} A(\bar{k},\omega,\text{T}) d \bar{k}$, where $\bar{k}$ defines the $\overline{\Gamma}$--$\overline{\text{Y}}$ direction) does not vary within a 3$\%$ uncertainty over the explored temperature range, consistent with general spectral-function sum rules (more details in Supplementary Materials).

In further support of our interpretation, we examine also the \text{T}-dependence of the spectral function at $\omega$=0, $A(0,\mathbf{k}_\mathrm{F},\text{T})$, which can be written as:
\begin{equation} \label{eq:3}
A(0,\mathbf{k}_\mathrm{F},\text{T})=
\begin{cases}
\frac{1}{\pi}\, \frac{1}{\Gamma+\beta(\pi k_B \text{T})^2} \,, & \text{for FL}\\
\frac{1}{\pi}\, \frac{1}{\Gamma+ \lambda(\frac{\pi}{2} k_B \text{T})} \,, & \text{for MFL}\\
\end{cases}
\end{equation}
where $\Gamma$ is a scattering rate term independent of energy and temperature as in Eq.\,\ref{eq:2}. We remark that the \text{T}-dependence in Eq.\,\ref{eq:3} is solely determined by the parameters $\beta$ and $\lambda$, offering a simpler and novel comparison with FL and MFL models. 
Figure\,\ref{Fig3}d compares the relative variation of the spectral function, $\Delta A(0,\mathbf{k}_\mathrm{F})$, for the two models of Eq.\,\ref{eq:3} (solid lines, normalized to the 10\,K value) to its experimental counterpart for Bi2201-OD24 and Bi2212-OP91. While the FL model captures remarkably well the experimental data, MFL instead fails in reproducing the observed T-dependence. This finding is consistent with optical and transport studies reporting evidences of a FL regime in the underdoped region of other high-T$_c$ cuprates \cite{mirzaei2013spectroscopic, barivsic2013universal, proust2016fermi}, as well as with a recent ARPES study of an overdoped La-based cuprate, which reports a FL-to-MFL crossover moving from the nodal to the antinodal region \cite{chang2013anisotropic}.

In conclusion, we have reported a comprehensive study of the meltdown of the nodal CSW as a function of the electronic temperature in Bi-based cuprates. By employing ARPES, both in its conventional and time-resolved fashion, we have investigated various doping levels and compounds over a broad temperature range, and revealed a ubiquitous T-dependence of the nodal CSW bearing no direct relation to the superconducting or pseuodgap temperature scales. Instead, the observed suppression of the nodal CSW at $\mathbf{k}_\mathrm{F}$ naturally stems from the temperature and energy dependence of the FL electron self energy. While our findings demonstrate the Fermi Liquid nature of quasiparticles along the gapless nodal direction of Bi-based cuprates, further investigations are needed to address whether such a scenario holds along the whole Fermi arc or if a crossover to a Marginal Fermi Liquid behavior occurs instead. To this end, the pump-probe approach presented here for the study of the CSW temperature evolution may soon be extended to the antinodal region of cuprates, owing to the advent of TR-ARPES in the extreme-ultra-violet regime \cite{mills2019cavity, na2019direct}.\\

\begin{acknowledgments}
We gratefully acknowledge E.\,Ostroumov for his technical support. This research was undertaken thanks in part to funding from the Max Planck-UBC-UTokyo Centre for Quantum Materials and the Canada First Research Excellence Fund, Quantum Materials and Future Technologies Program. This project is also funded by the Gordon and Betty Moore Foundation's EPiQS Initiative, Grant GBMF4779 to A.D. and D.J.J.; the Killam, Alfred P. Sloan, and Natural Sciences and Engineering Research Council of Canada's (NSERC's) Steacie Memorial Fellowships (A.D.); the Alexander von Humboldt Fellowship (A.D.); the Canada Research Chairs Program (A.D.); NSERC, Canada Foundation for Innovation (CFI); British Columbia Knowledge Development Fund (BCKDF); and the CIFAR Quantum Materials Program. E.R. acknowledges support from the Swiss National Science Foundation (SNSF) grant no. P300P2 164649. Part of the research described in this work was performed at the Canadian Light Source, a national research facility of the University of Saskatchewan, which is supported by the Canada Foundation for Innovation (CFI), the Natural Sciences and Engineering Research Council (NSERC), the National Research Council (NRC), the Canadian Institutes of Health Research (CIHR), the Government of Saskatchewan, and the University of Saskatchewan.
\end{acknowledgments}

\bibliographystyle{apsrev4-1}

\providecommand{\noopsort}[1]{}\providecommand{\singleletter}[1]{#1}

\end{document}